# Dynamics of large-scale solar-wind streams obtained by the double superposed epoch analysis. 3. Deflection of speed vector

Yu. I. Yermolaev, I. G. Lodkina, M. Yu. Yermolaev

**Abstract**

This work is a continuation of our previous articles (Yermolaev *et al.* in *J. Geophys.Res*. **120**, 7094, 2015; Yermolaev *et al.* in *Solar Phys.* **292**, 193, 2017), which describe the average temporal profiles of interplanetary plasma and field parameters in large-scale solar-wind (SW) streams: corotating interaction regions (CIRs), interplanetary coronal mass ejections (ICMEs including both magnetic clouds (MCs) and ejecta), and sheath as well as interplanetary shocks (ISs). Changes of longitude angle $\varphi$ in CIRs from -2 to +2° agree with earlier observations. Besides we have for the first time analyzed the average temporal profiles of bulk velocity angles in Sheaths and ICMEs and found that the angle $\varphi$ in ICME changes from 2 to −2° while in Sheath it changes from −2 to 2° (similar to change in CIR), *i.e.*, the streams in CIR/Sheath and ICME deflect in the opposite side. When averaging the latitude angle $\vartheta$ on all intervals of the chosen SW type, the angle $\vartheta$ is almost constant ~1°. We made selection of SW events with increasing and decreasing angle $\vartheta$ for the first time and found that average temporal profiles for angle $\vartheta$ in selected events have the same "integral-like" shape as for angle $\varphi$. The difference in average profiles for angles $\varphi$ and $\vartheta$ is explained by the fact that most events have increasing profiles for angle in ecliptic plane due to solar rotation while for angle in meridional plane numbers of increasing and decreasing profiles are equal.

## 1. Introduction

Quiet and uniform solar wind (SW) at 1 AU extends in the radial direction. Non-stationarity and heterogeneity in the solar atmosphere lead to formation of disturbed structures of solar wind which have different speeds and interact with surrounding quasi-stationary SW streams and with each other. One of the experimental evidence of SW streams interaction is the SW deflection relative to radial direction. Though the speed vector measured in separate events can change on an individual path, when averaging on a large number of events of identical SW type the change in the stream direction shows steady trends. For example, corotating interaction regions (CIRs) are a consequence of spatial variability in the coronal expansion and solar rotation which cause interaction of fast streams from coronal holes with slow streams from coronal belts. This interaction results in compression of plasma between fast and slow streams and deflection of SW speed vector from westward (negative longitude angle $\varphi$) to eastward (positive $\varphi$) direction (*e.g.*, see reviews by Gosling and Pizzo, 1999 and Tsurutani *et al.,* 2006 and references therein). Fast motion of interplanetary coronal mass ejections (ICMEs including flux rope magnetic clouds (MCs) and non-MC Ejecta) can interact with surrounding SW and form a compression region Sheath before ICME.

Recently we calculated the average temporal profiles of several interplanetary and magnetospheric parameters for eight usual sequences of SW phenomena: (1) SW/CIR/SW, (2) SW/IS/CIR/SW, (3) SW/Ejecta/SW, (4) SW/Sheath/Ejecta/SW, (5) SW/IS/Sheath/Ejecta/SW, (6) SW/MC/SW, (7) SW/Sheath/MC/SW, and (8) SW/IS/Sheath/MC/SW (where SW is undisturbed solar wind, IS means interplanetary shock) for interval 1976-2000 (Yermolaev *et al.,* 2015). To calculate the average temporal profile of parameters in the phenomena with different durations, we use the double method of superposed epoch analysis: rescaling the time between points of interval (proportionally increased/decreased) in such a way that respective beginnings and ends for all intervals of a selected SW type coincide (Yermolaev *et al.,* 2010, 2015). Figure 1 summarizes behavior of longitude $\varphi$ and latitude $\vartheta$ bulk velocity angles observed in 8 different sequences of phenomena. Changes of longitude angle φ in CIRs (both CIR sub-types: with IS and without IS) agree with earlier observations. Besides we have for the first time analyzed the average temporal profiles of bulk velocity angles in Sheaths and ICMEs and found that the angle φ in ICME changes from 2 to −2° while in Sheath it changes from −2 to 2° (similar to change in CIR), *i.e.,* the streams in CIR/Sheath and ICME deflect in the opposite sides (the angle φ shows large deflections in MC and Sheath before it because of small statistics). It has been explained by the interaction of fast ICME with slow plasma in SW before Sheath. Two types of compression regions CIRs and Sheaths are formed due to speed difference between undisturbed SW and leading edge of corresponding piston (High-Speed Stream (HSS) or ICME) and have identical average speed profiles in value and in inclination; the speed profiles for cases with an IS are located ~100 km s$^{-1}$ higher than for those without an IS, which indicates that the ~100 km s$^{-1}$ increase in speed of the pistons leads to the formation of an IS (Yermolaev *et al.,* 2017).

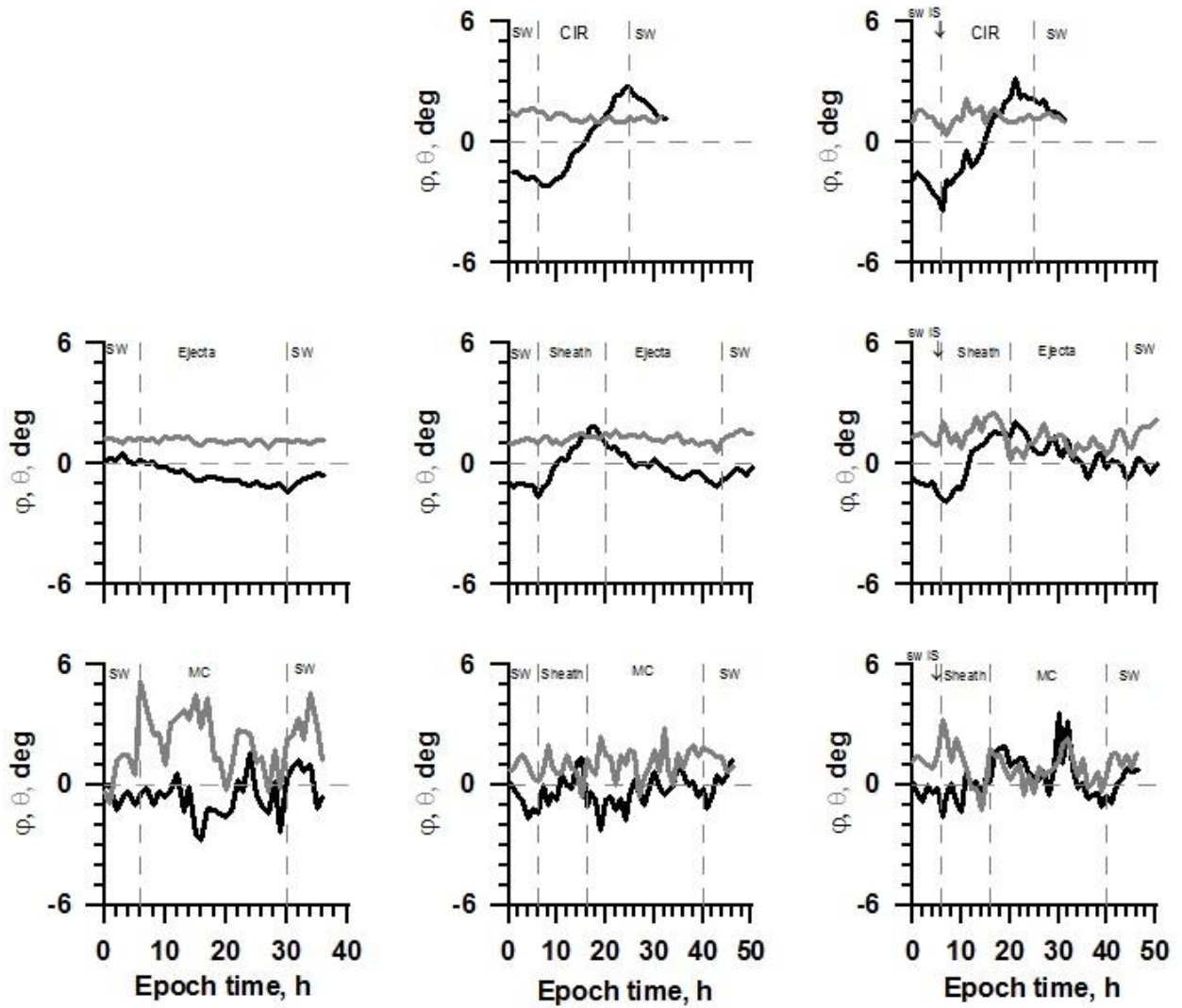

**Figure 1.** Time variation of longitude φ (black line) and latitude ϑ (grey) bulk velocity angles obtained by the double method of superposed epoch analysis in eight sequences of SW events: (1) CIR, (2) IS/CIR, (3) Ejecta, (4) Sheath/Ejecta, (5) IS/Sheath/Ejecta, (6) MC, (7) Sheath/MC, and (8) IS/Sheath/MC. Dashed vertical lines show boundaries of SW types which are indicated in the upper parts of each panel.

When averaging the latitude angle ϑ on all intervals of the chosen SW type, the angle ϑ is almost constant ~1°. We suggest that in contrast to the longitude angle φ which has one temporal profile for selected SW types, the latitude angle ϑ can have several different temporal profiles due to various direction of pistons motion relative to the equator plane and used procedure of data processing did not allow us to set off these profiles from total data set.

In the present work for the first time we make an additional selection on the basis of angle $\vartheta$ variation in the intervals to separate various profiles. The organization of the paper is as follows: section 2 describes data and method. In section 3, we present results on behavior of longitude $\varphi$ and latitude $\vartheta$ bulk velocity angles in various SW types. Section 4 summarizes the results.

## 2. Data and methods

In present paper we use the same data and methods as in our previous works (Yermolaev *et al.*, 2015; 2017): (1) the one hour (1-h) interplanetary plasma and magnetic field data of the OMNI database (http://omniweb.gsfc.nasa.gov (King and Papitashvili, 2004)), (2) our catalog of large-scale solar-wind phenomena during 1976-2000 (ftp://ftp.iki.rssi.ru/pub/omni/ (Yermolaev *et al.*, 2009)) and (3) the double superposed epoch analysis method (Yermolaev *et al.*, 2010). This method involves re-scaling (proportional increasing/decreasing time between points) the duration of the interval for all SW types in such a manner that, respectively, beginnings and ends of all intervals of a selected type coincide. Similar methods of profile analysis were used in the papers by Yokoyama and Kamide (1997); Lepping *et al.* (2003, 2017).

We use the same set of SW events with following small corrections: (1) we united data of two sets MC and Ejecta and analyze the total set ICME as the statistics of MCs is small, and (2) we excluded several events which have not measurements of angles during all intervals. As result our analysis includes 355 events for CIR, 350 for Sheath, and 357 for ICME.

We calculate values of angles $\varphi 1$ and $\varphi 2$, and $\vartheta 1$ and $\vartheta 2$ by averaging over first and second half of SW type interval, respectively, and select events with similar signs of corresponding angles, i.e. -/- (angles in both halves of interval are negative), -/+ (angle in first half of interval is negative and it in second half is positive), +/- and +/+. Such approach allows us to estimate approximately the main change (trend) of deflection of a stream in each event and to carry out selection of events with similar deflection. It should note that for the majority of data subsets the statistics of events remains enough large for reliable results.

## 3. Results

In this section, we perform two types of analysis of the speed angles $\varphi$ and $\vartheta$ in different types of solar wind. First, we study the trend of changing the angles, *i.e.* we investigate the relations of the longitude ($\varphi 1$ and $\varphi 2$) and latitude ($\vartheta 1$ and $\vartheta 2$) angles averaged, respectively, in the first and second halves of the intervals of the corresponding type of solar wind. In this case, the time resolution of the averaged angles is 5-12 hours and allows us to only roughly estimate the trend. Then, based on the obtained trend, we select the events and study the detailed (with a time resolution of about 1 hour) behavior of the average angles for events with a corresponding trend.

### 3.1. Analysis of trends of angles $\varphi$ and $\vartheta$

Figure 2 presents the angles $\varphi 1$ and $\varphi 2$ for 6 SW types: 1) CIR, (2) IS/CIR, (3) Sheath, (4) IS/Sheath, (5) ICME and (6) IS/ ICME. Numbers and percentages of events in 4 quadrants (-/+, +/-, -/-, and +/+) and in 2 half-planes (φ1>φ2 and φ1<φ2) of average angles φ1 and φ2 for 6 different SW types are presented in Table1. For CIRs, ~4/5 events have increasing angle φ (φ1<φ2) and ~1/2 events in -/+ quadrant. These numerous events result in the well-known average temporal profile of angle φ in CIRs (see top panels in Figure 1). Although there are fewer such events in Sheaths than in CIRs (~2/3 with φ1<φ2 and ~1/3 in -/+ quadrant, respectively), their relative number is sufficient to make the average profile of angle φ in Sheaths looks like the profile in CIRs. ICMEs have approximately the same number of events as Sheaths, but with the opposite trend: ~2/3 with φ1>φ2 and ~1/3 in +/- quadrant, respectively. Therefore, the opposite trend of the average angle profile is observed in ICMEs.

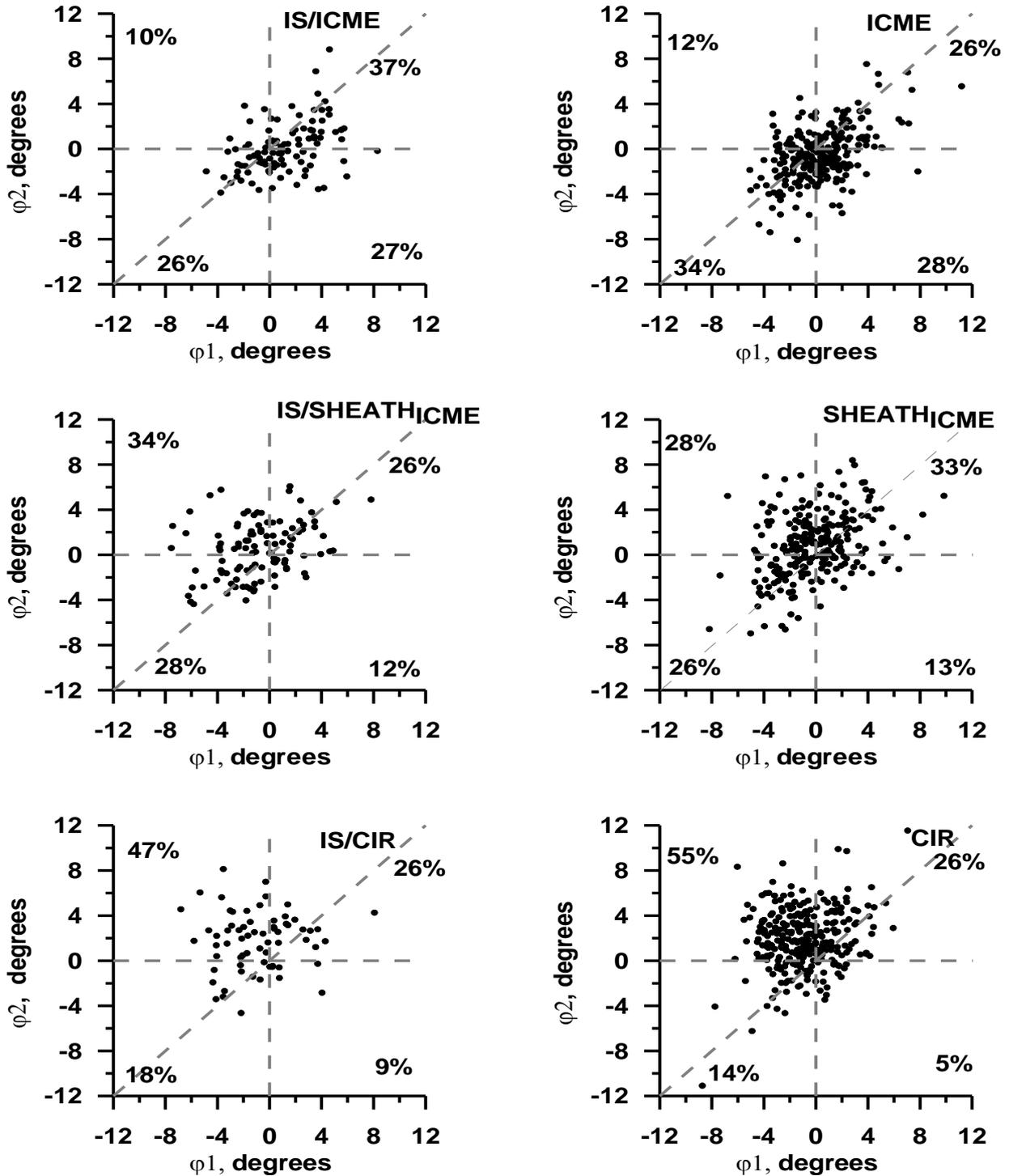

**Figure 2.** Distributions of average longitude velocity angles φ2 and φ1 for 6 different SW types indicated in right upper corners of panels. For each panel the percentages of events in 4 quadrants are presented.

**Table 1.** Number and percentage of events in 4 quadrants (-/+, +/-, -/-, and +/+) and in 2 half plane (φ1>φ2 and φ1<φ2) of average angles φ1 and φ2 for 6 different SW types (see Figure 2)

| SW type | Number of events (%) | | | | | | |
|---|---|---|---|---|---|---|---|
|  | All | φ -/+ | φ +/- | φ -/- | φ +/+ | φ1>φ2 | φ1<φ2 |
| CIR w/o IS | 289 | 159 (55) | 15 (5) | 40 (14) | 75 (26) | 53 (18) | 236 (82) |
| CIR with IS | 66 | 31 (47) | 6 (9) | 12 (18) | 17 (26) | 16 (24) | 50 (76) |
| Sheath$_{ICME}$ w/o IS | 243 | 69 (28) | 31 (13) | 62 (26) | 81 (33) | 85 (35) | 158 (65) |
| Sheath$_{ICME}$ with IS | 107 | 36 (34) | 13 (12) | 30 (28) | 28 (26) | 38 (36) | 69 (64) |
| ICME w/o IS | 260 | 33 (12) | 72 (28) | 88 (34) | 67 (26) | 162 (62) | 98 (38) |
| ICME with IS | 100 | 10 (10) | 27 (27) | 26 (26) | 37 (37) | 64 (64) | 36 (36) |

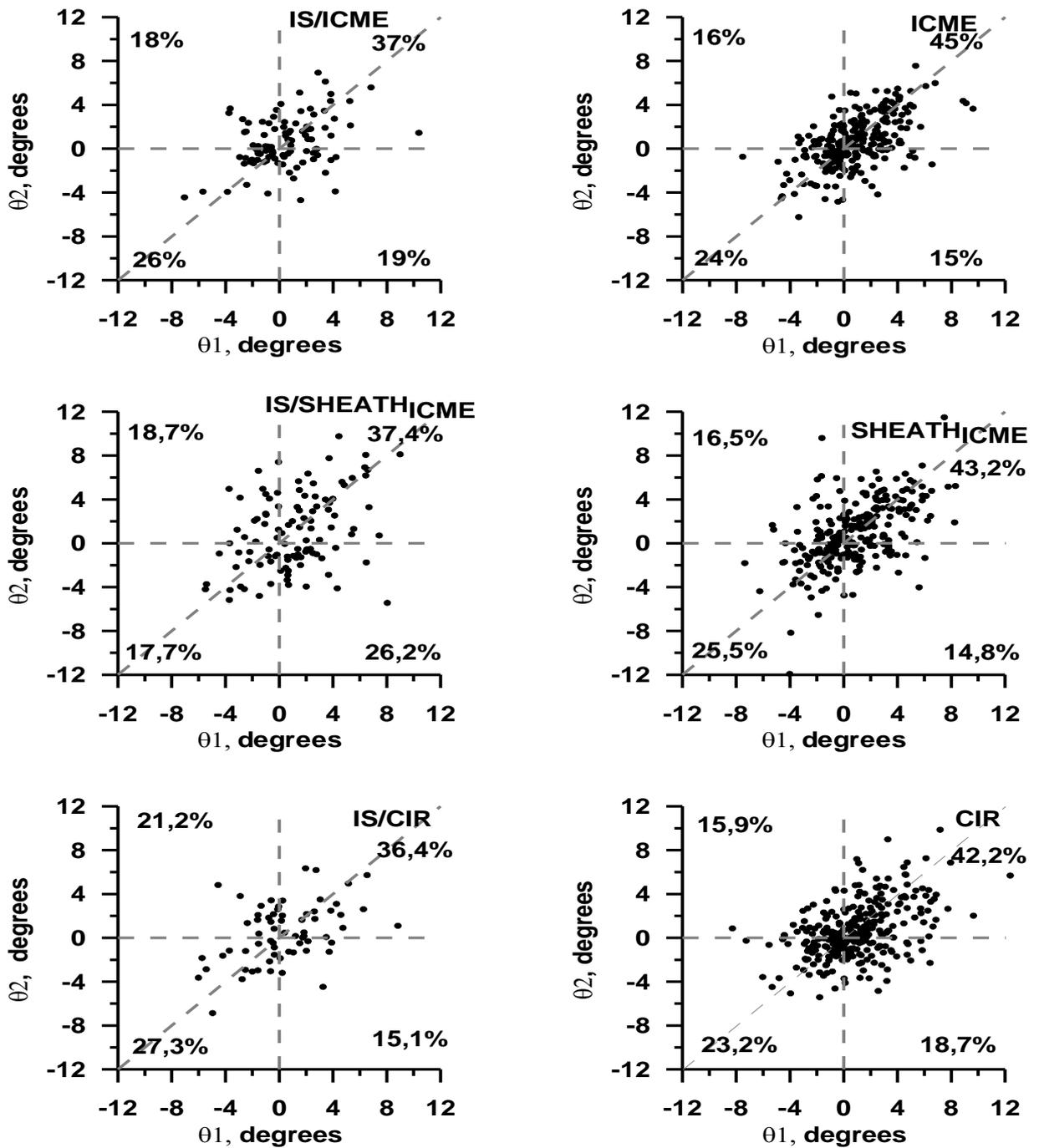

**Figure 3.** Distributions of average latitude velocity angles ϑ2 and ϑ1 for 6 different SW types

In contrast to the longitude angle φ, the latitude angle ϑ is fairly uniformly distributed over the quadrants and half-planes. Figure 3 presents average angles ϑ1 and ϑ2 for the same 6 SW types:

1) CIR, (2) IS/CIR, (3) Sheath, (4) IS/Sheath, (5) ICME and (6) IS/ ICME. Numbers and percentages of events in 4 quadrants and 2 half-planes are shown in Table 2. Main part of events for all SW types are located symmetrically near line $\vartheta_1 = \vartheta_2$ and in +/+ and -/- quadrants, *i.e.* ~60% events do not change the value and sign of angle $\vartheta$ during corresponding SW type interval.

Table 2. Number and percentage of events in 4 quadrants (-/+, +/-, -/-, and +/+) and in 2 half plane ($\vartheta_1 > \vartheta_2$ and $\vartheta_1 < \vartheta_2$) of average angles $\vartheta_1$ and $\vartheta_2$ for 6 different SW types (see Figure 3)

| SW type | Number of events (%) | | | | | | |
|---|---|---|---|---|---|---|---|
| | All | $\vartheta$ -/+ | $\vartheta$ +/- | $\vartheta$ -/- | $\vartheta$ +/+ | $\vartheta_1 > \vartheta_2$ | $\vartheta_1 < \vartheta_1$ |
| CIR w/o IS | 289 | 46 (16) | 54 (19) | 67 (23) | 122 (42) | 160 (55) | 129 (45) |
| CIR with IS | 66 | 14 (22) | 10 (15) | 18 (27) | 24 (36) | 34 (52) | 32 (48) |
| Sheath$_{ICME}$ w/o IS | 243 | 40 (17) | 36 (15) | 62 (25) | 105 (43) | 108 (44) | 135 (56) |
| Sheath$_{ICME}$ with IS | 107 | 20 (19) | 28 (26) | 19 (18) | 40 (37) | 52 (49) | 55 (51) |
| ICME w/o IS | 257 | 40 (16) | 39 (15) | 62 (24) | 116 (45) | 129 (50) | 128 (50) |
| ICME with IS | 100 | 18 (18) | 19 (19) | 26 (26) | 37 (37) | 41 (41) | 59 (59) |

Figure 4 shows a deflection of speed vector in longitude and latitude planes (changes of angles $\Delta\vartheta = \vartheta_2 - \vartheta_1$ and $\Delta\varphi = \varphi_2 - \varphi_1$) in different SW types. Average values, standard deviations and statistic errors of angle differences $\Delta\vartheta$ and $\Delta\varphi$ are presented in Table 3. Standard deviations for $\Delta\vartheta$ and $\Delta\varphi$ are close to each other and equal ~3$^o$. Owing to sufficiently high statistics, the mean values of the speed vector return in the ecliptic plane exceed statistical errors and indicate the presence of a systematic deflection in one direction for CIRs and Sheaths and the opposite direction for ICMEs. In meridianal plane, the mean angles of the speed vector return are less than statistical errors and their distributions are symmetric relative to radial direction.

Table 3. Average values and standard deviations (statistic errors) of angle differences $\Delta\vartheta$ and $\Delta\varphi$

| SW type | Number of events (N) | $\langle\Delta\vartheta\rangle \pm \sigma_{\Delta\vartheta}$ ($\sigma_{\Delta\vartheta}/\sqrt{N}$) | $\langle\Delta\varphi\rangle \pm \sigma_{\Delta\varphi}$ ($\sigma_{\Delta\varphi}/\sqrt{N}$) |
|---|---|---|---|
| CIR w/o IS | 289 | -0,26±2,83 (0,17) | 2,90±3,17 (0,19) |
| CIR with IS | 66 | 0,16±3,06 (0,38) | 2,74±3,73 (0,46) |
| Sheath$_{ICME}$ w/o IS | 243 | 0,14±2,83 (0,18) | 1,13±3,12 (0,30) |
| Sheath$_{ICME}$ with IS | 107 | -0,05±3,62 (0,35) | 1,43±3,14 (0,30) |
| ICME w/o IS | 257 | -0,12±2,56 (0,16) | -0,82±2,42 (0,15) |
| ICME with IS | 100 | 0,17±2,77 (0,28) | -0,96±2,67 (0,27) |

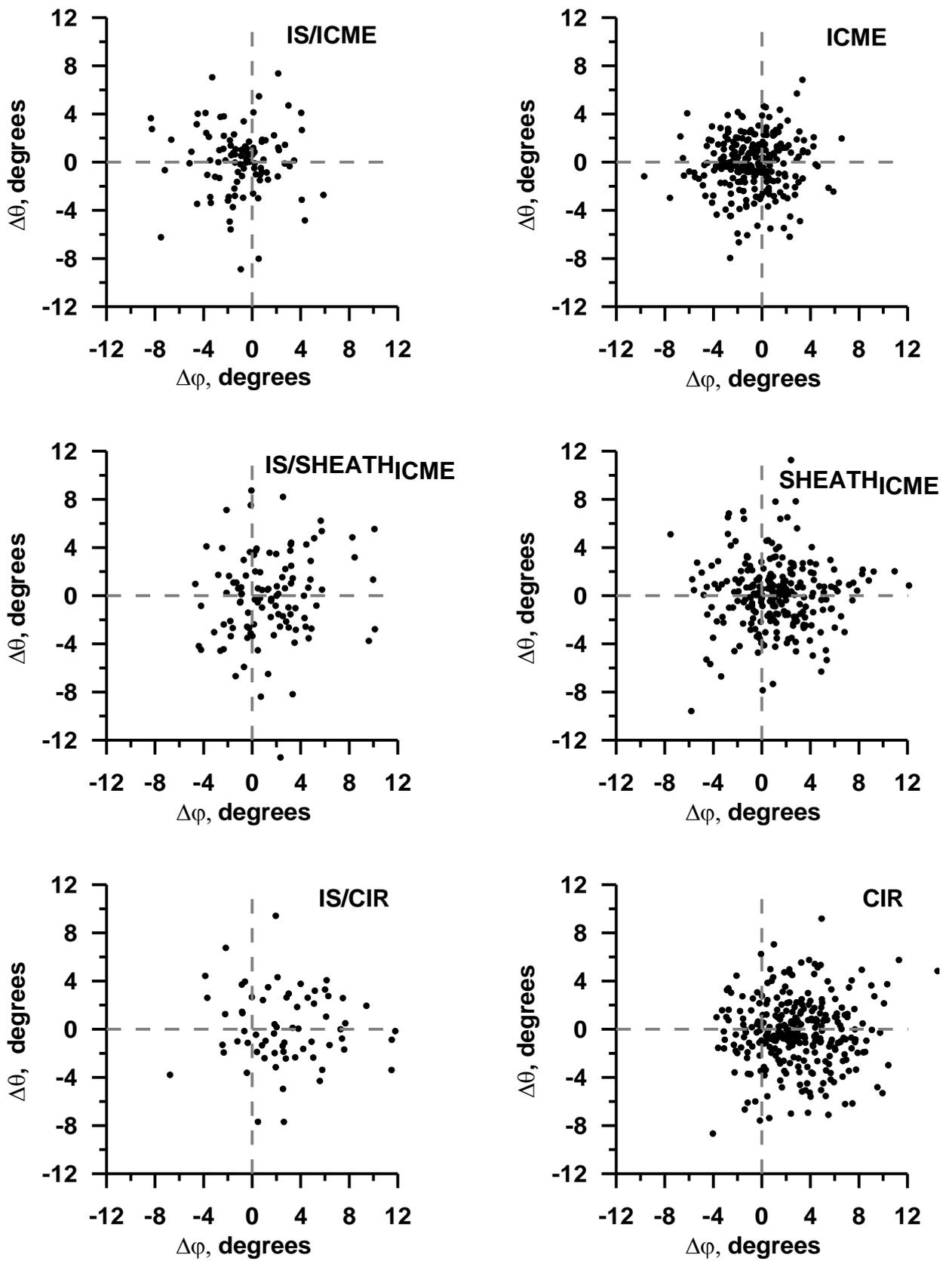

**Figure 4.** Distributions of differences of average longitude velocity angles $\Delta\varphi = \varphi2 - \varphi1$ and latitude velocity angles $\Delta\vartheta = \vartheta2 - \vartheta1$ for 6 different SW types

Figure 5 shows average direction of speed vector in first (red) and second (blue) halves of intervals for different SW types. Average values, standard deviations and statistic errors of angles $\vartheta$ and $\varphi$ in 1st and 2nd halves of intervals for different SW types are presented in Table 4. Data of Figure 5 and Table 4 agree with similar data of Figure 4 and Table 3 and in contrast with Figure 4 and Table 3 they give us additional information about start and end values of angles.

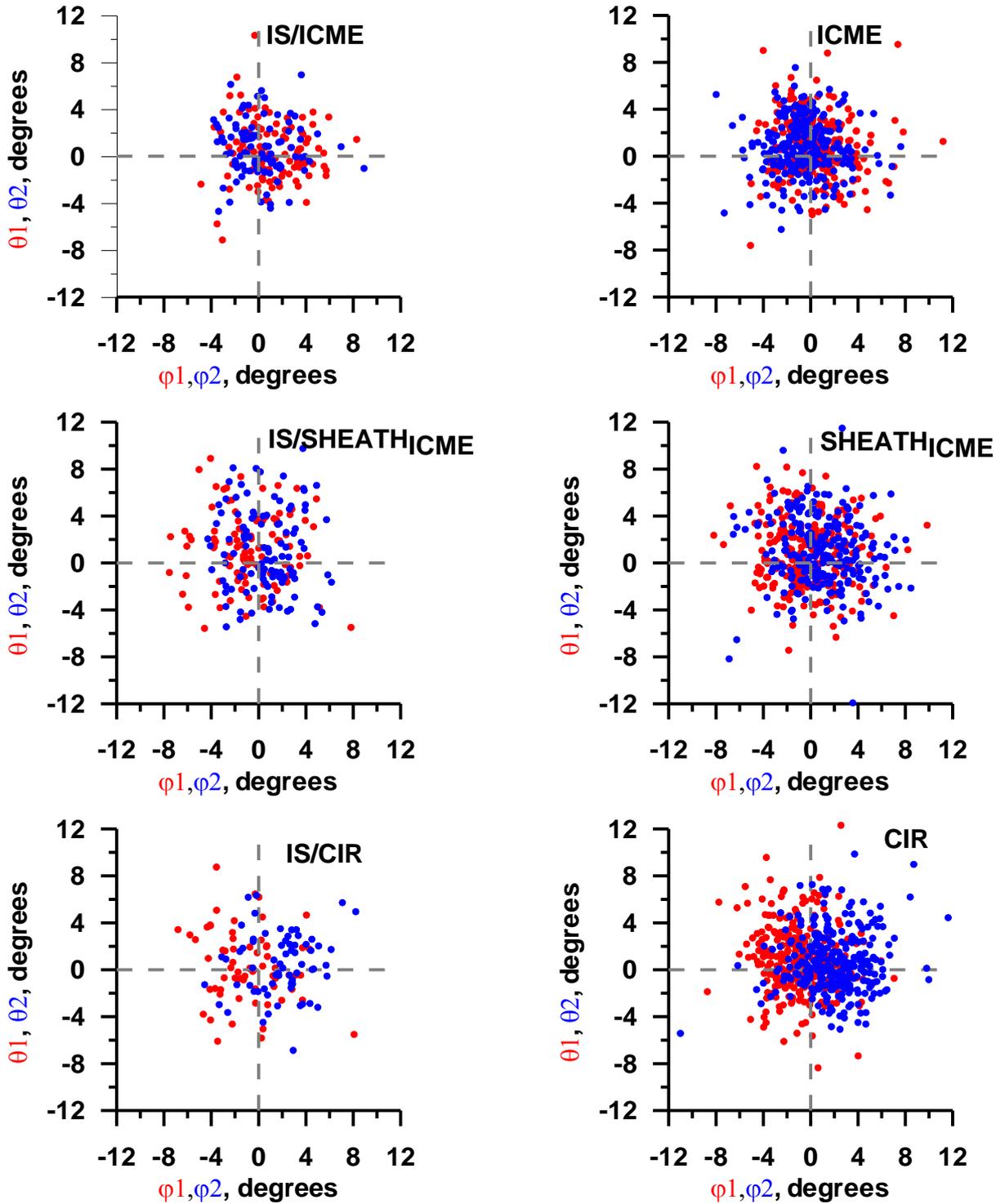

**Figure 5**. Stream directions in first halves of intervals (red circles) and in second halves of intervals (blue circles) for different SW types (ICME, SHEATH and CIR)

Table 4. Average values and standard deviations (statistic errors) of angles $\vartheta$ and $\varphi$ in 1st and 2nd halves of intervals for different SW types

| SW type | Number of events N | $<\vartheta_1> \pm \sigma_{\vartheta_1}$ ($\sigma_{\vartheta_1}/\sqrt{N}$) | $<\vartheta_2> \pm \sigma_{\vartheta_2}$ ($\sigma_{\vartheta_2}/\sqrt{N}$) | $<\varphi_1> \pm \sigma_{\varphi_1}$ ($\sigma_{\varphi_1}/\sqrt{N}$) | $<\varphi_2> \pm \sigma_{\varphi_2}$ ($\sigma_{\varphi_2}/\sqrt{N}$) |
|---|---|---|---|---|---|
| CIR w/o IS | 289 | 0,94±2,87 (0,17) | 0,68±2,52 (0,15) | -0,93±2,49 (0,15) | 1,94±2,65 (0,16) |
| CIR with IS | 66 | 0,35±3,02 (0,37) | 0,51±2,66 (0,33) | -0,95±2,78 (0,34) | 1,79±2,58 (0,32) |
| SHE$_{ICME}$ w/o IS | 243 | 0,78±2,82 (0,18) | 0,92±2,92 (0,19) | -0,24±2,71 (0,17) | 0,90±2,81 (0,18) |
| SHE$_{ICME}$ with IS | 107 | 1,17±3,03 (0,29) | 1,11±3,46 (0,33) | -0,89±2,90 (0,28) | 0,54±2,45 (0,24) |
| ICME w/o IS | 257 | 0,84±2,52 (0,16) | 0,71±2,31 (0,14) | 0,28±2,45 (0,15) | -0,51±2,36 (0,15) |
| ICME with IS | 100 | 0,46±2,61 (0,26) | 0,63±2,32 (0,23) | 1,10±2,58 (0,26) | 0,14±2,28 (0,23) |

### 3.2. Analysis of temporal profiles of angles $\varphi$ and $\vartheta$

In this subsection we analyze the average temporal profiles of angles $\varphi$ and $\vartheta$ in different SW types when selecting the events with 4 trends of angles $\vartheta$: -/+, +/-, -/-, and +/+. It is necessary to indicate that the event selection results in significant decrease in statistics and several profiles are not sufficiently smooth. Nevertheless the obtained data allow us to make conclusions on main features of profiles.

Figure 6 shows results obtained by double superposed epoch analysis for CIRs. In all subsets the angle $\varphi$ has similar "integral-like" shape and does not depend on the selection with angle $\vartheta$. During increasing and decreasing trends of angle $\vartheta$ (-/+ and +/-) the average temporal profiles of angle $\vartheta$ show corresponding profiles with similar shapes: first the modulus of the angle increases, reaches a maximum, decreases to 0 and then has an analogous form with the opposite sign. For cases without trends (without a change of sign +/+ and -/-), the form for $\vartheta$ has, respectively, a maximum or minimum in the middle of the interval.

Figure 7 shows results obtained for the complex of events "Sheaths + ICMEs". First and second columns of the figure present results when selecting the events in Sheaths, and third and fourth columns in ICMEs. Behavior of the angle $\varphi$ are similar to one before $\vartheta$ selection: in all subsets the angle $\varphi$ has similar "integral-like" shape and does not depend on selection with angle $\vartheta$. The average temporal profiles of angle $\vartheta$ in Sheaths and ICMEs behave the same ways as in CIRs in accordance with the trends -/+ or +/-. It is important to note that values of angle $\vartheta$ in end of Sheath corresponds to one in beginning of ICME: there are no discontinuations and events with increasing (decreasing) trend in Sheaths have the similar shape for events with decreasing (increasing) trend in ICMEs (see, for example, pairs Fig.7b/Fig.7k and Fig.7c/Fig.7i).

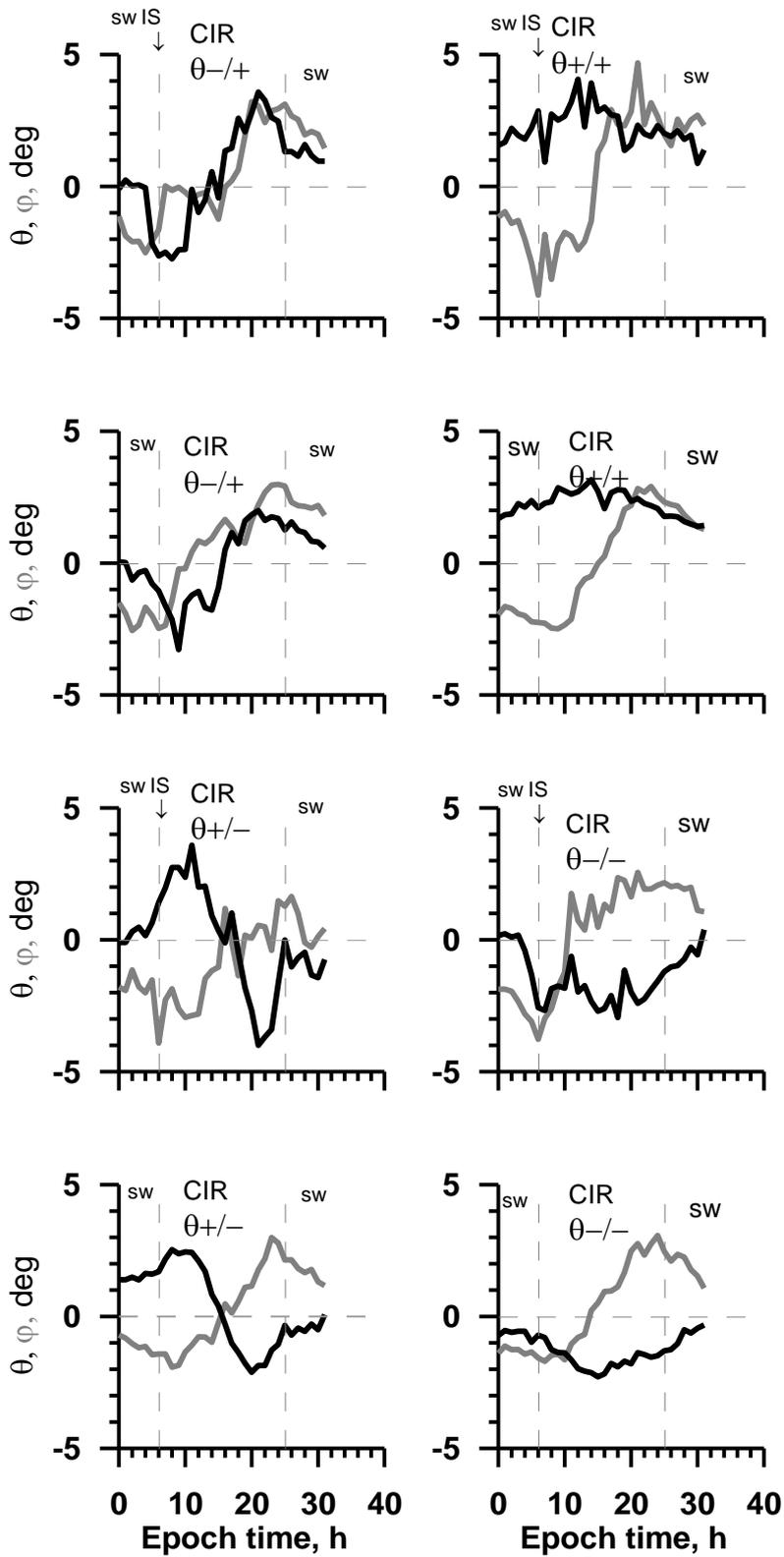

**Figure 7**. Average temporal profiles of velocity angles φ and ϑ for CIRs when selecting the events with trend of ϑ -/+, -/-, +/+, and -/+.

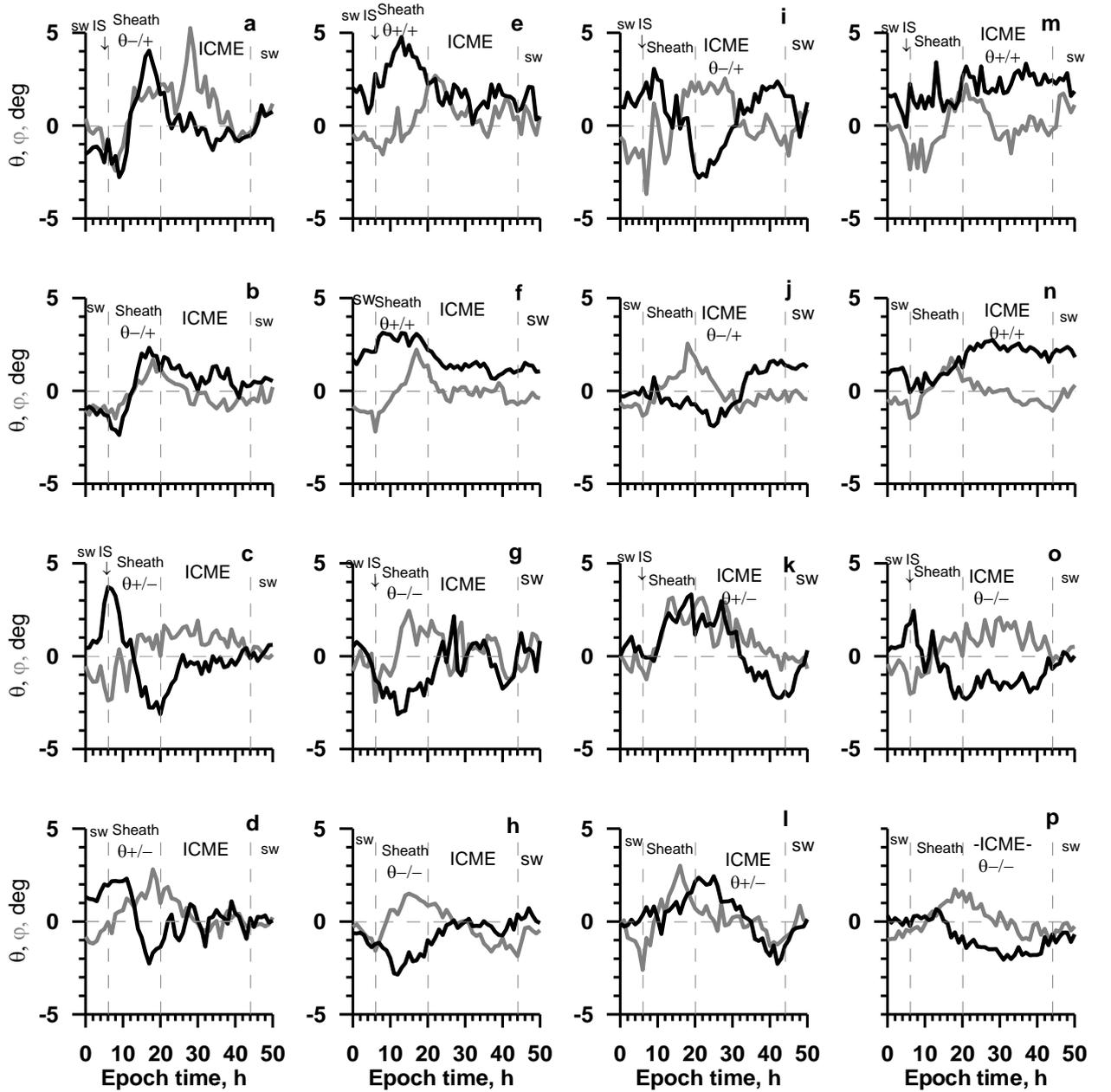

**Figure 7**. Average temporal profiles of speed angles φ and ϑ for Sheaths and ICMEs when selecting the trend of ϑ -/+, -/-, +/+, and -/+ for Sheaths (1-st and 2-nd columns) and ICMEs (3-rd and 4-th columns).

## 4. Discussion and Conclusions

Thus we statistically studied the variation of longitude and latitude speed angles φ and ϑ in CIRs, Sheaths and ICMEs (both with and without interplanetary shocks before CIRs and Sheaths) and obtained the following results.

1. For CIR and Sheath, the numbers of events with an increasing trend in the longitude angle φ (φ1 < φ2) are about 80 and 65%, respectively, and the average time variation of the angle φ has an increasing "integral-like" form. For ICME, the number of events with a decreasing trend in the angle φ is about 63%, and the average time variation of the angle φ has a decreasing shape.

2. For all SW types, the number of events with an increasing and decreasing trends in the latitude angle ϑ is small and approximately equal to each other (~15%), the number with a positive angle (+ / +) is slightly higher than the number with a negative angle (- / -) (40% vs. 25%), and therefore the average time variation of the angle ϑ is approximately constant and is about 1°.

3. The presence of interplanetary shocks before CIRs and Sheaths has little effect on the above distribution of events.

4. For all cases of selection with respect to $\vartheta$, the average angle $\varphi$ has the same shape in CIRs and Sheaths, i.e. it does not depend on selection by $\vartheta$; in ICME the average profile of angle $\varphi$ has a irregular shape (due to insufficiently high statistic) and tends to either decrease or be constant.

5. For Sheaths and CIRs with the increasing and decreasing trends of the angle $\vartheta$ (with a change of sign of trends: - / + and +/-), the form for $\vartheta$ is the same as for $\varphi$: first the modulus of the angle increases, reaches a maximum, decreases to 0 and then has the analogous form with the opposite sign. For cases without trends (without a change of sign of trends: + / + and - / -), the form for $\vartheta$ has, respectively, a maximum or minimum in the middle of the interval.

6. For all cases of selection with respect to $\vartheta$ in Sheath, the mean values of angle $\vartheta$ in ICME are small and vary little over the interval. The values of $\vartheta$ at the end of the Sheath and the beginning of the ICME are close to each other.

7. For all cases of selection with respect to $\vartheta$ in the ICME, the mean values of angle $\vartheta$ in Sheath are small and vary little over the interval. The values of $\vartheta$ at the end of Sheath and the beginning of ICME are close to each other.

The obtained results show that, regardless of the type of piston, the interaction of large-scale solar wind streams leads to the same scenario of the direction changes of the speed vector in longitudinal and latitudinal planes in accordance with the physical conditions in the interaction region. Observation of the angles deflection can be considered as a marker of these interactions.

**Acknowledgements** We thank the OMNI database team for the opportunity to use data obtained from GSFC/SPDF OMNIWeb (http://omniweb.gsfc.nasa.gov). YY is grateful to the SCOSTEP "Variability of the Sun and Its Terrestrial Impact" (VarSITI) program for support of his participation in first General Symposium (VarSITI2016) in Albena, Bulgaria, 6-10 June, 2016. This work was supported by the Russian Science Foundation, project 16-12-10062.